\newcommand{\HI}{H\,\textsc{i}\xspace}
\newcommand{\kms}{~km\,s$^{-1}$\xspace}

\documentclass{aa}

\usepackage{graphicx}
\usepackage{txfonts}
\usepackage{xcolor}
\usepackage{multirow}
\usepackage{threeparttable}
\begin{document} 

   \title{A depolarizing \HI tidal tail in the western lobe of Fornax\,A}
   
   \author{F. Loi\inst{1} \and P. Serra\inst{1} \and
          M. Murgia\inst{1} \and F. Govoni\inst{1} \and 
          C. Anderson\inst{2} \and G. Heald\inst{3} \and D. Kleiner\inst{1} 
          \and E. Lenc\inst{4} \and
          V. Vacca\inst{1} \and
          F. M. Maccagni\inst{1,5} \and
          R.~J. Dettmar\inst{6}.
          }
    \authorrunning{F. Loi \and et al.}
    \titlerunning{A depolarizing \HI tidal tail in the western lobe of Fornax\,A}

   \institute{INAF-Osservatorio Astronomico di Cagliari, via della Scienza 5, 09047 Selargius\\
        \email{francesca.loi@inaf.it}
        \and National Radio Astronomy Observatory, 1003 Lopezville Rd, Socorro, NM 87801
        \and CSIRO, Space and Astronomy, PO Box 1130, Bentley WA 6102, Australia
        \and CSIRO, Space and Astronomy, PO Box 76, Epping, NSW 1710, Australia
        \and Netherlands Institute for Radio Astronomy, Oude Hoogeveensedijk 4, 7991 PD Dwingeloo, The Netherlands
        \and Ruhr University Bochum, Faculty of Physics and Astronomy, Astronomical Institute, 44780 Bochum, Germany}
   \date{Received December 10, 2021; accepted ****, ****}

  \abstract
    {Recent MeerKAT neutral hydrogen (\HI) observations of Fornax\,A reveal tidal material intersecting in projection the western lobe of this radio galaxy. We found a spatial coincidence between the northern \HI tail and a depolarized structure observed for the first time with the Australian Square Kilometre Array Pathfinder (ASKAP) at 1.2 GHz. We analyzed the properties of the rotation measure (RM) image obtained with ASKAP data at the location of the \HI tail and in its neighborhood.
    We modeled the observed RM structure function to investigate the magnetic field power spectrum at the location of the \HI tail and in a nearby control region. We found that the observed RM, in the control region and in a region enclosing the \HI tail, cannot be due to the intracluster Faraday screen caused by the Fornax cluster. An intragroup magnetized medium with a central magnetic field strength of 18.5\,$\rm\muup$G can explain the control region RM, but it is clear that there is an excess in correspondence with the \HI tail region. We evaluated several scenarios in which the \HI tail is either in the lobe foreground or embedded in the lobe. We determined a magnetic field strength on the order of $\sim$9.5$-$11 $\muup$G in the \HI tail, a value consistent with constraints derived from  narrowband H$\alpha$ imaging of the ionized gas. The spatial coincidence between \HI tail and depolarization analyzed in this paper could be the first observed evidence of a magnetic field that either has passed through a radio galaxy lobe or has survived the lobe expansion.}

   \keywords{Galaxies: individual: Fornax A -- galaxies: magnetic fields -- galaxies: ISM -- galaxies: intergalactic medium -- polarization -- radio lines: galaxies}

   \maketitle
%

\section{Introduction}
In the last decade broadband spectropolarimetric observations have shown a pronounced network of patches and/or filamentary structures associated with the lobes of some radio galaxies \citep{brienza,fanaroff21,maccagni,ramatsoku,anderson18a,anderson18b}, both in total intensity and polarization.
Detailed studies of these systems also reveal the presence of  ``low-p'' patches, which are  regions with  low total intensity and polarization emission, depolarization structures, and rotation measure (RM) enhancements \citep{anderson18b,guidetti11,guidetti12}. Even though the depolarization observed for radio galaxies has usually been ascribed to foreground Faraday screens \citep{dreher,laing,guidetti10}, such as the intracluster medium of clusters, it is clear now that the interaction between the lobes and the environment can play a key role. Shock waves \citep{carilli,hardcastle12,banfield}, hydrodynamical or magneto-hydrodynamical instabilities, and shocked interstellar medium \citep{hardcastle} can cause the observed depolarized structures and RM enhancements.\\
Fornax\,A is a spectacular example of a radio galaxy showing lobes with a filamentary structure and depolarized patches. Thanks to its proximity \citep[$\rm D_L$ = 20.8 $\pm$ 0.5 Mpc;][]{cantiello}, these features were already evident based on the VLA data published by \citet{fomalont}. 
The radio galaxy is the brightest member of a galaxy group located southwest of the center of the Fornax galaxy cluster, at a projected separation of $\sim$ 1.3 Mpc, and it is likely falling towards the cluster central regions \citep{drinkwater}.
The radio lobes of Fornax\,A were generated during several intermittent episodes of active galactic nucleus (AGN) activity, as recently demonstrated by \citet{maccagni}, and extend over a linear size of $\sim$290 kpc. 
The two lobes do not show hot spots and are not connected to the AGN with visible jets \citep[S-shaped radio jets are observed within 6 kpc from the host galaxy center;][]{geld1,geld2,maccagni}. Therefore, the source cannot be classified as a Fanaroff-Riley type I or type II radio galaxy.\\
\citet{anderson18b} study in detail the polarization properties of the two lobes using the Australia Telescope Compact Array (ATCA) between 1.28 and 3.1 GHz with a resolution of 20\arcsec$\times$30\arcsec. These images show several low-p patches across the eastern lobe from the northeast to southwest, and several regions in the western lobe as well. Here in particular, two low-p patches correspond to regions where the depolarization is due to known foreground objects:  NGC\,1310 \citep{fomalont,schulman} and the Ant, which is associated with an extragalactic cloud of ionized hydrogen with  a recessional velocity similar to that of  Fornax\,A \citep{bland}. In the remaining regions, line-of-sight magnetic field reversals have been observed, and therefore \citet{anderson18b} interpret these low-p patches as being due to the interaction with the environment: magnetized thermal plasma is advected from the interstellar medium of NGC\,1316, from the peculiar early-type galaxy hosting Fornax\,A, or from the intracluster or intragroup medium along the line of sight (or close to it) generating RM gradients and low emission because of the magnetic field orientation.\\
The galaxy NGC\,1316 is the result of a 10:1 merger that occurred $\sim$2\,Gyr ago between a massive early-type galaxy and a gas-rich late-type galaxy. During the merger, stellar and gaseous tails and loops formed and are now visible in the outskirts of the stellar body \citep{schweizer,mackie,goud,serra}.
The \HI content of NGC\,1316 and its group have been studied by several authors. Very recently, \citet{kleiner}  observed the NGC\,1316 group with the MeerKAT radio telescope reaching a column density sensitivity of 1.4 $\times 10^{19}$ atoms cm$^{-2}$. These authors estimate that, during the merger, 7$-$11$\cdot$10$^8$ M$_{\odot}$ of \HI was ejected to a large radius by tidal forces, and is now detected as clouds and tails in the intragroup medium.

\begin{figure*}
\centering
\includegraphics[width=0.99\textwidth]{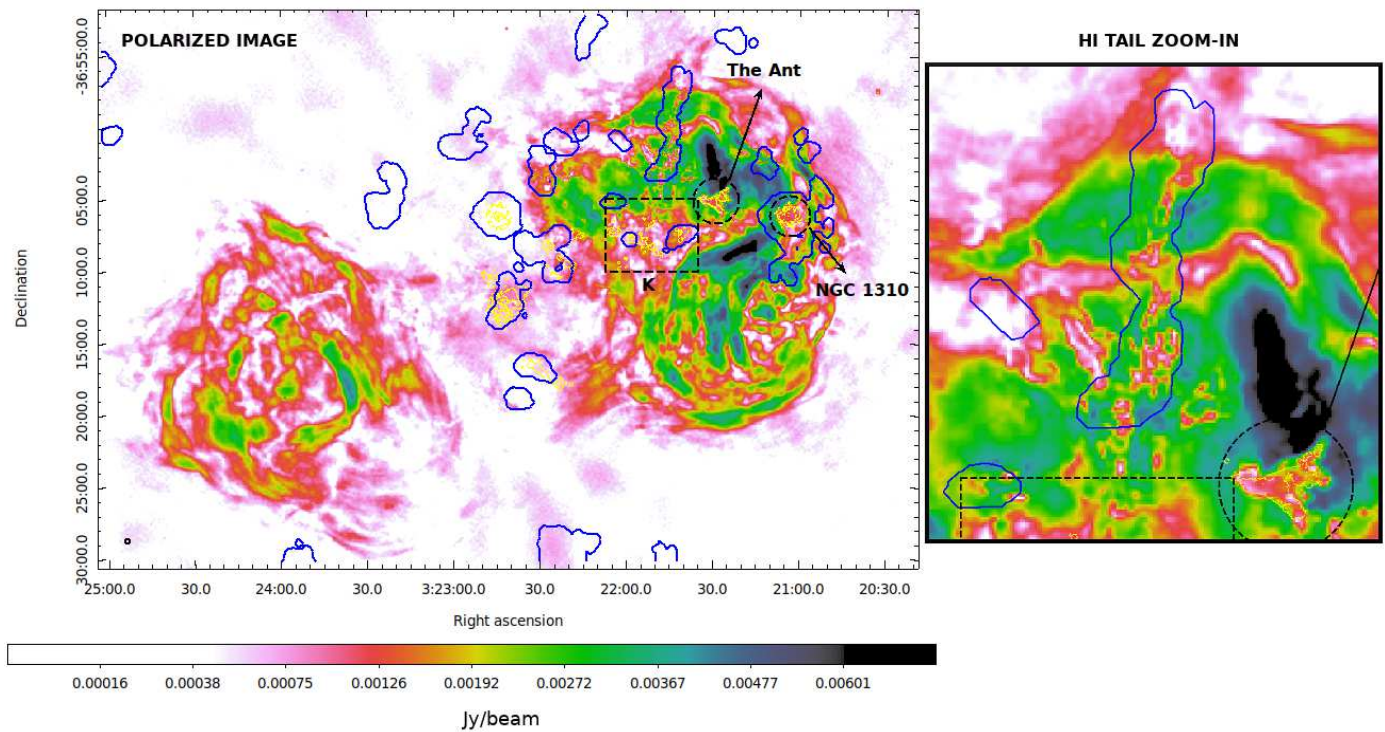}
\caption{ASKAP polarized image (defined as the peak of the Faraday dispersion function) and, on the right, a zoomed image centered on the southern part of the \HI tail. The beam size is shown in the  bottom left corner of the image; its FWHM corresponds to 12\arcsec. In both images VST H$\alpha$ and MeerKAT \HI contours are overlaid in yellow and blue, respectively. The H$\alpha$ contour starts at 2$\rm \cdot 10^{-18} erg \cdot cm^{-2} \cdot s^{-1} \cdot arcsec^{-2}$, while the \HI contour represent a column density of N$_{\HI}$=1.4$\times$10$^{19}$ atoms\,cm$^{-2}$. The black dashed regions and arrows indicates the position of NGC\,1310, of the Ant, and the region labeled ``K'' in \citet{anderson18b}.}
\label{fig:pol}
\end{figure*}
\begin{figure*}
\centering
\includegraphics[width=0.99\textwidth]{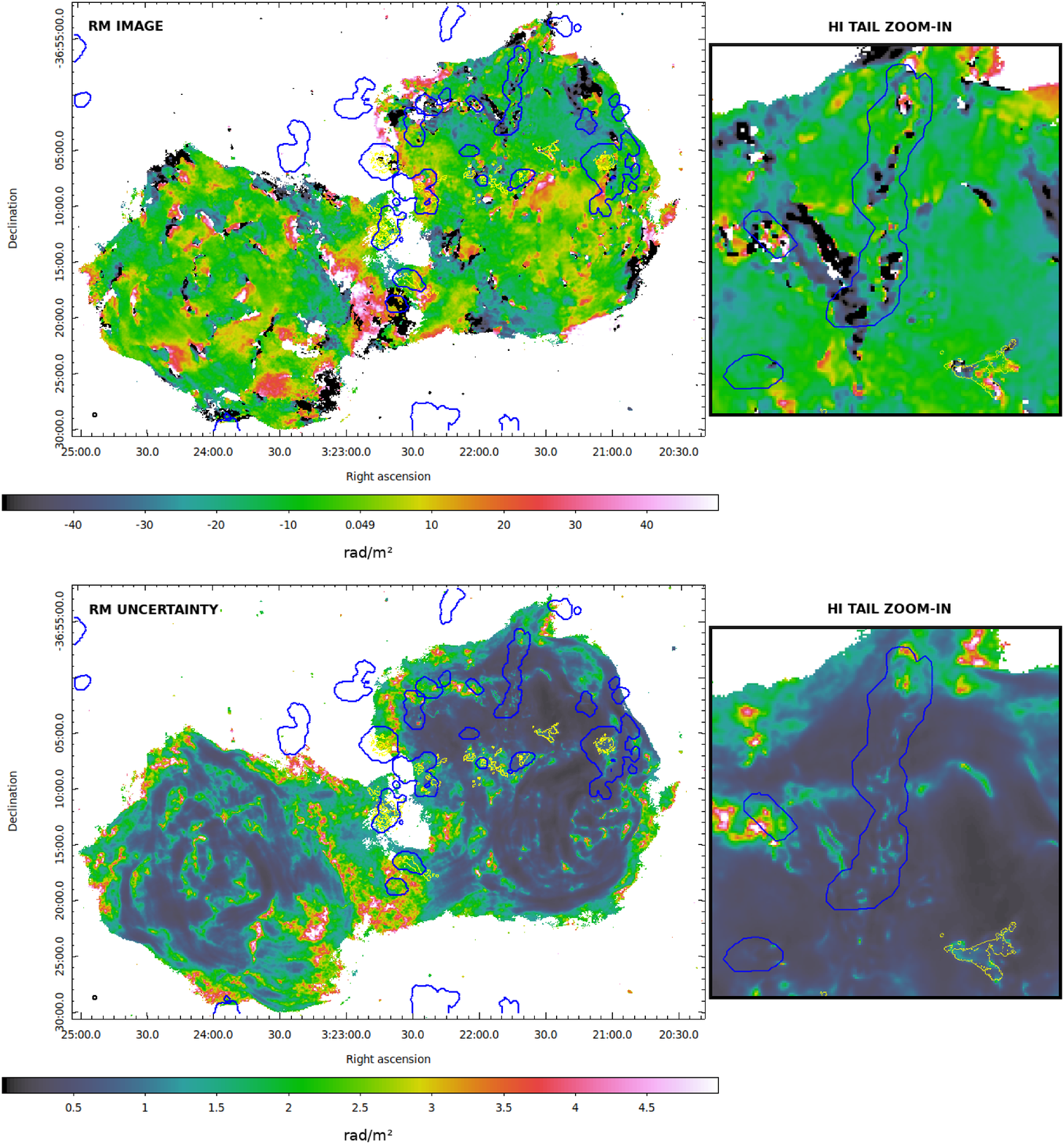}
\caption{ASKAP RM image (top), defined as the Faraday depth at the peak of the Faraday dispersion function, and its uncertainty (bottom). The beam size is shown in the  bottom left  corner of the image; its FWHM corresponds to 12\arcsec. In both images VST H$\alpha$ and MeerKAT \HI contours are overlaid  in yellow and  blue, respectively. The H$\alpha$ contour starts at 2$\rm \cdot 10^{-18} erg \cdot cm^{-2} \cdot s^{-1} \cdot arcsec^{-2}$, while the \HI contour represent a column density of N$_{\rm HI}$=1.4$\times$10$^{19}$ atoms\,cm$^{-2}$.}
\label{fig:fd}
\end{figure*}
In this work we present the first observed spatial coincidence between neutral hydrogen tidal material detected by MeerKAT and a depolarization pattern observed with the Australian Square Kilometre Array Pathfinder (ASKAP) in the western lobe of Fornax\,A. 
We analyze the properties of the RM along the \HI tail and in a similar region to characterize the observed association.\\ 
In Section 2 we present the data used throughout the paper; in Section 3 we determine the RM properties in the \HI tail region and in a control region with the same dimensions and orientation; in Section 4 we present the 2D numerical simulations performed in order to determine the power spectrum properties of the magnetic fields; in Section 5 we use 3D numerical simulations  to interpret the data; and in Section 6 we use H$\alpha$ observations to obtain upper limits on the thermal plasma density. Finally, we summarize the results and draw our conclusions in Section 7.
Throughout this paper we assume a luminosity distance of (20.8$\pm$0.5)\,Mpc to NGC\,1316 \citep{cantiello,hatt}. At this distance, 1 arcmin corresponds to 5.8 kpc.
   
\section{Data}
The data analyzed in this work consists of the {\HI} image obtained with MeerKAT and already published by \citet{kleiner}, together with the polarized image and RM image obtained with ASKAP during commissioning tests of the telescope in view of the Early Science program which covered the Fornax cluster \citep{anderson21}. The commissioning observations presented here correspond to a different scheduling block with respect to the Fornax cluster observations by \cite{anderson21}. As we describe in the following, the settings are not the same. Nevertheless, the data reduction and the imaging were carried out in a very similar way. Here, we briefly describe the observational setup, the data reduction, and the imaging of these data sets.

\subsection{MeerKAT \HI data}
The MeerKAT \HI images were produced from two commissioning observations conducted in June 2018. The observations were both performed with the SKARAB correlator in 4k mode, which consists of 4096 channels in full polarization in the frequency range 856-1712\,MHz with a resolution of 209\,kHz (equivalent to 44.1 \kms for \HI at the distance of the Fornax cluster). The two observations had 36 and 62 antennas connected to the correlator; the target was observed for 8\,h and 7\,h, respectively. 
The full data reduction process is described in \citet{kleiner} and we refer to that paper for further details.

Relevant to this work is the Northern \HI tail (T$_{\rm N}$) that overlaps in projection with the Fornax\,A western lobe. T$_{\rm N}$ was first detected by \citet{serra}, where they showed that NGC\,1316 underwent a 10:1 merger 1 -- 3\,Gyr ago, and tidal forces ejected \HI in the intragroup medium out to large radii ($\sim$ 150 kpc)  in the form of two tails.  The northern \HI tail is made of tidally stripped material once located within the spiral galaxy involved in the merger that formed NGC\,1316. \citet{kleiner} detected additional \HI components of the \HI tails that doubled the mass and length of T$_{\rm N}$. In this work we explore how the \HI of T$_{\rm N}$ may be interacting with the Fornax\,A western lobe and influencing its polarization properties.  

\subsection{ASKAP polarized data and RM}
We observed Fornax\,A during commissioning tests of the ASKAP radio telescope \citep{DeBoer2009, Johnston2007, schinckel2016} on April 3, 2019. We observed a single pointing (SBID 8377, start time 2019-04-03 01:48:39.435877) for ten hours using a single formed beam \citep{McConnell2016}, covering a 288 MHz frequency range spanning 800$-$1088 MHz, achieving a measured band-averaged sensitivity of 24 $\muup$Jy beam$^{-1}$ at 10{\arcsec} per Stokes parameter, and recording the full set of polarization products. \\
The data reduction and imaging were performed as described in \cite{anderson21} and we refer to that work for more details.
The only difference with respect to that work consists in the spatial resolution used to smooth the final images (12\arcsec$\times$12\arcsec), which is the spatial resolution of our lowest frequency channel. As in \cite{anderson21}, the Q and U datacubes with dimensions RA, Dec, $\rm\lambda^2$ were used to compute the Faraday Dispersion Spectrum (FDS) over the range from -200 to +200 rad m$^{-2}$ using RM synthesis\footnote{\url{https://github.com/brentjens/rm-synthesis}, version 1.0-rc4} \citep{Burn1966,B2005}. The result is a complex-valued FDS datacube with dimensions RA, Dec, and $\phi$. We generated maps of the peak polarized intensity (peak-$P$) and Faraday depth at the peak of the FDS (hereafter RM, for historical reasons) across the field from the FDS cube using {\tt Miriad}'s {\sc moment} function \citep{Sault1995}.

Figure \ref{fig:pol} shows the polarized intensity of Fornax\,A with blue and yellow contours respectively representing the MeerKAT \HI emission and the VLT Survey Telescope (VST) H$\alpha$ emission mapped with narrowband deep photometric observations taken with the VST-OmegaCAM \citep{kleiner}.
The Fornax\,A western lobe has a filamentary shape with the well-known low-p patches \citep{fomalont,anderson18b} and new features revealed by the sensitivity and resolution of the ASKAP image.
The zoomed-in image in the right panel  of Figure \ref{fig:pol} shows the \HI tail.
We note that the polarization properties are significantly different along the \HI tail with respect to close-by regions. In particular, we can see the presence of a small-scale ($\leq 2$ kpc) depolarized structure along the \HI tail. This pattern seems to extend beyond the tail towards the south, connecting the \HI tail with the low-p patch labeled ``K'' in \citet{anderson18b}, which we indicate with a dashed box in the figure. The region called the  Ant is also highlighted in Figure \ref{fig:pol} (dashed circle). Here, as reported in the Introduction, a cloud of ionized hydrogen has been detected and it is acting as a Faraday screen.
\begin{figure*}
   \resizebox{\hsize}{!}
   {\includegraphics[width=\textwidth]{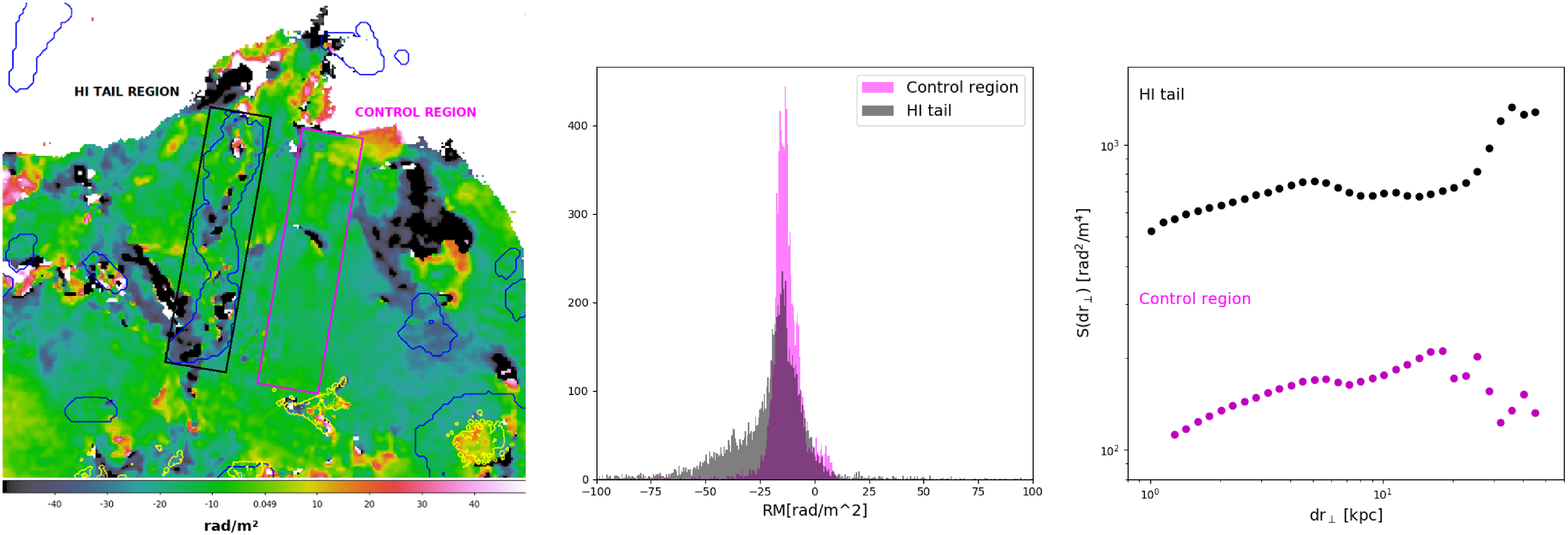}}
      \caption{RM images and properties of the \HI tail and in a nearby region. {\it Left}: Magnified RM image with VST H$\alpha$ contours (yellow) and MeerKAT \HI contours (blue). The black and magenta boxes indicate the \HI tail and control regions considered in our analysis. {\it Middle}: Distribution of the RM values in the \HI tail (black) and in the control region (magenta). {\it Right}: RM structure functions evaluated in the \HI tail (black) and in the control region (magenta).}
    \label{fig:rm_box}
\end{figure*}

The RM image is shown in Figure \ref{fig:fd} (top image). We subtracted from the image a constant value of 6 rad/m$^2$ due to the Galactic foreground. Figure \ref{fig:fd} also shows the RM uncertainty (bottom image), computed according to Eq. 13 in \citet{sotomayor}:
\begin{equation}
    \rm \Delta \phi (l) = \frac{1}{S/N} \frac{\sqrt{3}}{\lambda_2^2-\lambda_1^2},
\end{equation}
where S/N is the signal-to-noise ratio, while $\rm\lambda_1$ and $\rm\lambda_2$ are the minimum and maximum wavelength of the bandwidth.
The RM image presents regions with a high RM scatter. These are found in correspondence of the low-p patches, as already pointed out by \citet{anderson18b}. However, also in this case, we observe a new feature along the \HI tail which had not been detected probably because of the very high resolution achieved with the ASKAP image. Here we can see an enhancement of the RM scatter on small scales. Similar patterns can be observed in other locations in correspondence of either \HI emission or H$\alpha$ detections. In the next section we characterize the RM in the tail region with the aim of understanding whether the properties along the \HI tail are different with respect to the nearby regions. For the sake of simplicity, in this paper we focus only on the \HI tail since the tidal nature of this emission is well understood compared with the less certain origin of the nearby \HI clouds \citep{kleiner}.

\section{Characteristic of the RM in the \HI tail region}
We compared the properties of the RM in a rectangular region which encloses the \HI emission and in a nearby box with the same dimensions and orientation. We chose this control region according to two criteria. First, it has to be as close as possible to the \HI tail to meet similar physical conditions; second, it should not contain detected \HI or H$\alpha$ emission. We want to understand the role of the \HI tail as depolarizing screen, and need to exclude any additional components when comparing it with the lobe.\\
The two regions are shown in the left image of Figure \ref{fig:rm_box} as black and magenta boxes for the \HI tail and the control region, respectively. In the central panel of this figure, we also show the distribution of the RM values in the two regions.
The mean, median, and the standard deviation of the RM are equal to -20\,rad/m$^2$, -17\,rad/m$^2$, and 20\,rad/m$^2$ for the \HI tail
region and -13\,rad/m$^2$, -13\,rad/m$^2$, and 9\,rad/m$^2$ for the control. The average RM over the entire western lobe is equal to -7 rad/m$^2$. We recall that the galactic RM has been subtracted from the RM image.\\
We also evaluated the RM structure function which is defined as the average mean square difference in RM values between pixels at a given distance:
\begin{equation}
\rm    S(dr)=<[RM(r)-RM(r+dr)]^2>.
\end{equation}
The uncertainty associated with the RM structure function has been computed by propagating the RM errors:
\begin{equation}
    \rm \Delta S(dr)=\frac{2\sqrt{2S(dr)}}{\sqrt{N}} \Delta \phi (l).
\end{equation}

The results are shown in the right plot of Figure \ref{fig:rm_box} for the \HI tail (in black) and for the control region (in magenta). The error bars are within the points.\\
It is clear that the two RM structure functions have a different normalization. As more clearly seen in the next section, the shape and the slope are also different in the two regions. Assuming a uniform density of the thermal gas, this likely indicates a different geometry and strength of the underlying magnetic fields.

\section{Modeling the magnetic field power spectrum}
Numerical simulations can be used to constrain the properties of magnetic fields starting from spectropolarimetric radio observations. Using the FARADAY tool \citep{murgia}, we want to determine the power spectrum properties of the magnetic field associated with the observed RM and its strength with 2D and 3D simulations, respectively.\\
The RM power spectrum is proportional to the magnetic field power spectrum. Assuming a particular magnetic field power spectrum model, we can therefore derive the RM power spectrum and compute the RM structure function. By comparing the model RM structure function with the observed one, we can determine the magnetic field power spectrum repeating this procedure until we find the best-fit model. To reach our goal we used the least squares method.
We followed this procedure assuming a random Gaussian magnetic field power spectrum characterized by a minimum and a maximum scale of fluctuations, namely $\rm \Lambda_{min}$ and $\rm \Lambda_{max}$, a normalization and a slope n: $\rm |B_k|^2 \propto k^{-n}$. The magnetic field power spectrum presents degeneracy between its parameters. We decided to fix the minimum and maximum scale of fluctuations to 0.6 kpc and 46 kpc respectively for both the \HI tail region and the control region. Such scales correspond to half of the beam size of the RM image and to the length of the box used to evaluate the RM structure functions, respectively. We simulated the noise in the model images by adding a Gaussian noise with rms equals to the maximum uncertainty in the \HI tail and control region, which corresponds to 4\,rad/m$^2$. 
We performed 50 realizations of each different combination of slope and normalization of the magnetic field power spectrum, varying the slope between 1 and 2.4 with a step of 0.2, and the normalization from 50 to 500\,$\muup$G Mpc$^3$ in steps of 25\,$\muup$G Mpc$^3$.
\begin{figure}
    \centering
    \includegraphics[width=0.49\textwidth]{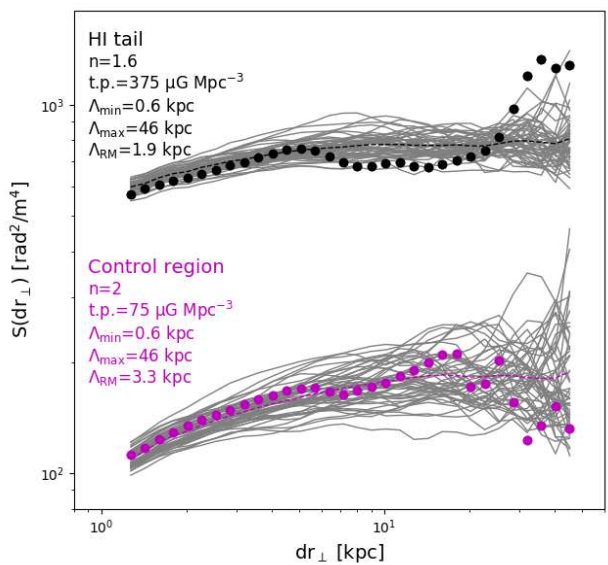}
    \caption{RM structure functions evaluated in the \HI tail (black) and in the control region (magenta). The gray lines are, for each RM structure function, 50 simulated functions generated from magnetic field power-law power spectrum models with parameters reported on the left.}
    \label{fig:2d}
\end{figure}

Figure \ref{fig:2d} shows the results of our analysis.
The gray lines are respectively the 50 different realizations of the best-fit magnetic field power spectra with parameters reported in the top and middle left corner for the \HI tail and the control region, respectively. The dashed black and magenta lines represent the average simulated RM structure function for the \HI tail and control region, respectively. We found a magnetic field power spectrum with a slope of n=1.6 and a total power of 375\,$\muup$G Mpc$^3$ in the \HI tail region, and a slope of n=2 and a total power of 75\,$\muup$G Mpc$^3$ in the control region. The RM auto-correlation length is $\rm\Lambda_{RM}$=1.9\,kpc and 3.3\,kpc for the \HI tail and the control region, respectively. With this power spectra we obtain magnetic fields with auto-correlation length $\rm\Lambda_c$ equal to 1.3\,kpc and 1.9\,kpc for the \HI tail and the control region, respectively.\\
The different properties of the two magnetic field power spectra would suggest that the two regions host different magnetic fields under the assumption of a uniform density of the thermal gas.

\section{RM interpretation}
In the following paragraphs we discuss and analyze several Faraday screen models which are due to the intracluster medium of the Fornax galaxy cluster (Sect. 5.1), to the intragroup medium of the Fornax\,A group (Sect. 5.2), and to a Faraday screen generated in the lobe surface and within the \HI tail (Sect. 5.3). We also investigate the possibility of a fluctuating density of the thermal plasma within the \HI tail (Sect. 5.4) and the physical implications of the \HI tail being inside the lobe (Sect. 5.5). These models are used to find the best representation of the data in the control region and in the \HI tail region too. \\
To reach our goal we first evaluated the magnetic field strength needed to cause the observed depolarization with a simple analytic approach. This gives us an order of magnitude estimate which we improved with 3D numerical simulation: starting from the magnetic field power spectrum models, we can assume a magnetic field and a thermal plasma distribution and derive the corresponding RM image. Again, we can compute the RM structure function from the RM simulated image and we can in principle determine the structure and the strength of the magnetic field by changing the model parameters and comparing with the observed RM structure function. It is worth noting that also in this case there is degeneracy between parameters. In particular, the RM depends on the magnetic field strength and morphology, on the thermal plasma density, and on the integral path length. The choice of the value of these parameters strictly depends on the interpretation of the data.

\subsection{Faraday screen from the Fornax intracluster medium}
We derived an estimate of the RM at the western lobe location assuming a magnetic field tangled on a single scale $\rm\Lambda_c$, randomly oriented from cell to cell \citep[see Appendix A of][]{loi19}. The result is a Gaussian-like distribution of the RM with zero mean and dispersion $\rm\sigma_{RM}$ \citep{lawler,felten}. Assuming a $\rm\beta$-model profile for the thermal plasma
\begin{equation}
   \rm n=n_0 \left( 1 + \left( \frac{r}{r_c}\right)^2\right)^{-3\beta/2},
\end{equation}
with central density $\rm n_0$ and core radius $\rm r_c$, and a magnetic field strength radial profile which follows the thermal plasma profile
\begin{equation}
    \rm B=B_0 \left( \frac{n}{n_0} \right)^{\eta},
    \label{eq:b}
\end{equation}
the RM standard deviation is a function of the projected distance according to \citet{dolag}:
\begin{equation}
    \rm   \sigma_{RM}(r)=K B_0 \Lambda_c^{0.5} n_0 r_c^{\frac{1}{2}} \frac{1}{(1+\frac{r^2}{r_c^2})^{\frac{6\beta(1+\eta)-1}{4}}} \sqrt{ \frac{\Gamma[3\beta(1+\eta)-\frac{1}{2}]}{\Gamma[3\beta(1+\eta)]}}.
   \label{eq:fel}
\end{equation}
The factor K is a constant that depends on the integral path that we consider to go from the Fornax\,A surface lobe to a distance of 1\,Mpc. We assumed the Fornax\,A western lobe to be spherical with a radius of 80\,kpc and that its center and the Fornax cluster center are at the same distance from us (i.e., they lie on the same plane perpendicular to the line of sight). Therefore, with the cluster center as the origin of our reference frame, we started the integral path from 80\,kpc. While we do not know the characteristics of the Fornax intracluster magnetic field, the thermal plasma distribution was constrained by \citet{chen} assuming the beta profile with a core radius of $\rm r_c$=173 kpc, $\rm \beta$=0.804, and a central density $\rm n_0=9\cdot 10^{-4}$ cm$^{-3}$. 
With these values at the Fornax\,A location (i.e., at a distance of r=1.3\,Mpc from the cluster center) we should observe RM values on the order of $\rm\sigma_{RM}\approx$0.01$\rm\cdot B_0$\,rad/m$^2$ assuming a $\rm\Lambda_c$=1.98\,kpc (see Section 2) and an index $\rm\eta$=0.5 \citep[see][]{govoni}. In the framework of this model to observe a $\rm\sigma_{RM}=9\,rad/m^2$, as in the control region, we would need a central magnetic field strength of 900\,$\muup$G. It is clear that the RM at the Fornax\,A location cannot be due to the intracluster medium of the Fornax cluster since magnetic field strengths in clusters are at most on the order of tens of $\muup$G \citep{vacca12}.

\subsection{Faraday screen from the Fornax\,A intragroup}
We considered the RM due to the Fornax\,A group to explain the observed RM in the control region.
We assumed the $\rm\beta$-profile for the thermal plasma by \citet{babyk} with $r_c$=0.21 kpc, $\rm\beta$=0.43, and n$_0$=11.76\,cm$^{-3}$ (adopting a mean mass of 0.63 times the proton mass).
Using Eq. \ref{eq:fel}, an integral path starting from the lobe surface up to 1\,Mpc of distance, $\rm\eta$=0.5, and $\rm\Lambda_c$=1.98\,kpc, we found $\rm\sigma_{RM}\approx$0.3$\cdot$ B${_0}$\,rad/m$^2$.
To explain the observed $\rm\sigma_{RM}$ in the control region, where this quantity equals 9\,rad/m$^2$, we would need a central magnetic field strength of B$_0\approx$30\,$\muup$G.\\ 
We ran 3D numerical simulations to obtain a better estimate of the intragroup magnetic field from the RM structure functions. In this case we created a simulated RM image starting from the best-fit magnetic field power spectra described in the previous section, which determine the geometry of the fields. We then assumed the $\rm\beta$-model by \citet{babyk}, a magnetic field profile following the plasma density distribution with $\rm\eta$=0.5 \citep{govoni}, and we performed several simulations changing the central magnetic field strength between 1 and 60\,$\muup$G with steps of 0.5\,$\muup$G to vary the RM normalization. For each value of the magnetic field strength, we performed 50 realizations. We determined B$_0$ with the least squares method and this resulted in 50 and 18.5\,$\muup$G in the case of the \HI tail and control region, respectively. 

\begin{figure}
    \centering
    \includegraphics[width=0.49\textwidth]{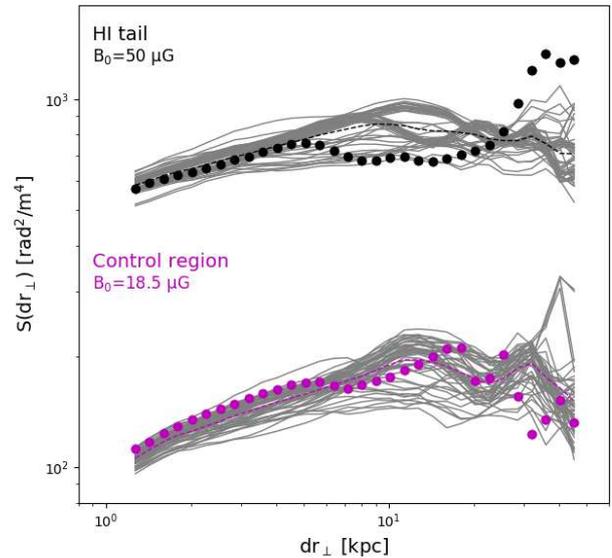}
    \caption{RM structure functions evaluated in the \HI tail (black) and in the control region (magenta). The gray lines are, for each RM structure, 50 simulated functions generated assuming a Faraday screen due to the Fornax\,A intragroup with magnetic field central strength reported on the left.}
    \label{fig:sdr_igm}
\end{figure}

The results are shown in Figure \ref{fig:sdr_igm}.
The RM structure function model for the control region follows the observed trend, while the adopted modeling cannot trace the behavior of the \HI region and in particular the decrease starting at 5\,kpc and then the rapid increment at $\sim$25\,kpc.
The hypothesis of a RM due to an intragroup Faraday screen cannot explain simultaneously the RM observed in the \HI tail and in the control region. In other words, the variation of the thermal plasma density from the \HI tail to the control region predicted by \citet{babyk} is insufficient to obtain consistent values for the central magnetic field strength. This means that at the \HI tail region we are detecting an additional Faraday screen whose nature could be due to the \HI tidal material itself.\\

Considering that $\rm\sigma_{RM}$=9\,rad/m$^2$ in the control region, to reach $\rm\sigma_{RM}$=20\,rad/m$^2$ at the \HI location we would need an extra $\rm\sigma_{RM}$ of 18\,rad/m$^2$ from the tail itself. 
Assuming a single-scale magnetic field within the \HI tail, B$_{\rm HT}$, which extends for a length L=11.04\,kpc, namely the box width of Figure \ref{fig:rm_box}, we know that this contribution can be parametrized as 
\begin{equation}
    \rm \sigma_{RM}=812\cdot \frac{B_{HT}}{\sqrt{3}}n_e \sqrt{\Lambda_C} \sqrt{L},
\end{equation}
where n$_e$ is the thermal density within the \HI tail, and $\rm\Lambda_c$ is the magnetic field auto-correlation length. According to the modeling of the magnetic field power spectra in Section 4, for the \HI region $\rm\Lambda_c$ is equal to 1.30\,kpc.
Using this formula, we would need a magnetic field of
$\rm B_{HT}\approx(0.010/n_e)\,\muup G$. 
If we consider a thermal density similar to the Fornax\,A value, which is on the order of 4$\cdot$10$^{-3}$cm$^{-3}$, B$_{\rm HT}\approx$2.5\,$\muup$G in the \HI tail.
The Fornax\,A intragroup medium modeling suggests a magnetic field of $\sim$0.35\,$\muup$G at the control region position. \\
Whatever the magnetized plasma generating the RM observed in the western lobe of Fornax\,A, it seems clear that there is an additional Faraday screen in correspondence of the \HI tail.

\subsection{A Fornax\,A shell lobe and a Faraday screen \HI tail}
As suggested by \citet{anderson18b}, the RM gradients observed in the low-p patches of the Fornax\,A lobes could be due to the advection of magnetized thermal plasma on the lobe surface. We can therefore explore a simple scenario in which the observed RM in correspondence of the control region is entirely due to a magnetized plasma on the surface of the lobe. The thermal plasma density is assumed to be constant, while the magnetic field fluctuates according to the models in Section 4 with a given average strength B$_0$. 

\begin{figure}
    \centering
    \includegraphics[width=0.49\textwidth]{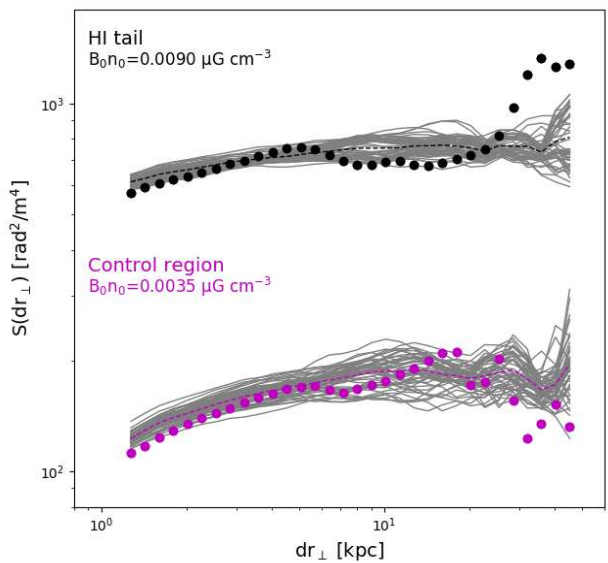}
    \caption{RM structure functions evaluated in the \HI tail (black) and in the control region (magenta). The gray lines are best-fit simulated RM structures obtained integrating over a path length of 11.04\,kpc assuming constant thermal plasma densities n$_0$ and  magnetic fields showing fluctuations around an average strength B$_0$. The best-fit parameters are reported on the left.}
    \label{fig:sdr_double}
\end{figure}

We started simulating the RM structure function due to the lobe surface. As already mentioned, the Faraday RM shows degeneracy in three parameters: the magnetic field strength, the thermal plasma density, and the path length. This means that we cannot derive an estimate for the magnetic field at the control region, but what is important here is to model the lobe RM in order to derive information about the \HI tail magnetic field.
We considered a path length of 11.04 kpc, which is the width of the control region, and we assume that the thermal plasma density is constant within this path length, while the magnetic field fluctuates around a given value B$_0$. Assuming this picture, we explored the parameter space between B$_0\cdot$n$_0$=1$\cdot10^{-3}\,\muup$G\,cm$^{-3}$ and B$_0\cdot$n$_0$=60$\cdot 10^{-3}\,\muup$G\,cm$^{-3}$ in steps of B$_0\cdot$n$_0$=0.5$\cdot10^{-3}\,\muup$G\,cm$^{-3}$. By minimizing the $\rm\chi^2$, we determined B$_0\cdot$n$_0$=3.5$\cdot 10^{-3}\,\muup$G\,cm$^{-3}$ for the control region, as can be seen in Figure \ref{fig:sdr_double}.
This modeling also seems to reproduce quite well the observed RM structure function.\\
If the emission detected in the direction of the \HI tail region is outside of the lobe in the foreground, then the observed RM is the sum of the Faraday rotation due to the lobe surface (RM$\rm_{lobe}$) plus the Faraday rotation due to the \HI tail (RM$_{\rm tail}$). The Faraday dispersion function, in the absence of Faraday complexity within the two contributions, will show a single polarized peak.
If, on the contrary, the \HI tail is within the lobe then we expect to find two Faraday components in the Faraday dispersion function: the polarized signals emitted at different path lengths in the lobe will suffer from the Faraday rotation due to the lobe surface (RM$\rm_{lobe}$) or a combination of this term plus the one due to the \HI tail (RM$_{\rm lobe}$+RM$_{\rm tail}$). It is clear that this scenario will lead to a Faraday complexity in the direction of the \HI tail region, an aspect that has been already evidenced by \citet{anderson18b}, even if it is not clear whether this involves only regions where ionized and/or molecular gas have been detected. Assuming that the Faraday complexity consists of two polarized peaks crossing the two different Faraday screens (i.e., in the absence of internal depolarization either within the lobe surface or in the \HI tail), we might wonder which polarized peak and corresponding RM were selected to create the RM image (which technically corresponds to the maximum polarized peak in the Faraday dispersion function): is it the component due to the sum RM$_{\rm lobe}$+RM$_{\rm tail}$ or simply to the RM$\rm_{lobe}$?
Since the RM at the \HI location is dramatically different with respect to the control region RM, we think that in this case the observed RM at the \HI tail region is the sum of RM$_{\rm lobe}$+RM$_{\rm tail}$.\\
Keeping all of these considerations in mind, we simulated two different scenarios in the \HI tail direction.
We first assumed that the RM is entirely due to the \HI tail Faraday screen. This will lead to an upper limit on the magnetic field estimate since we are assuming that only this field is responsible for the Faraday rotation. In this case we obtained B$_0 \cdot$n$_0$=9$\cdot 10^{-3}$\,$\muup$G cm$^{-3}$ as shown in Figure \ref{fig:sdr_double}.
\begin{figure}
    \centering
    \includegraphics[width=0.49\textwidth]{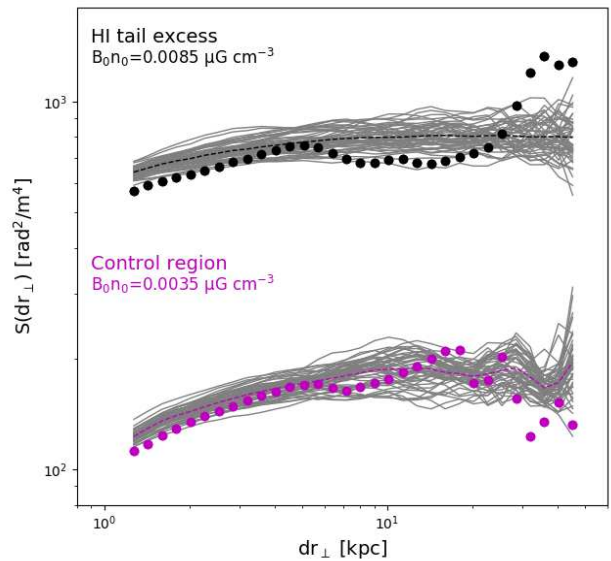}
    \caption{RM structure functions evaluated in the \HI tail (black) and in the control region (magenta). The gray lines near the control region are the best-fit simulated RM structures reported in Figure \ref{fig:sdr_double}. Those near the \HI tail region are best-fit models for the \HI tail excess obtained considering that the observed RM at the \HI position is due to the intrinsic \HI tail contribution plus the RM model of the control region. The best-fit parameters are reported on the left.}
    \label{fig:sdr_sum}
\end{figure}

The second scenario considers the Faraday rotation in the \HI tail region as the sum of the contributions of the \HI tail itself and the Fornax\,A lobe surface that we already modeled. In this scenario the \HI tail could be either in the lobe foreground or right behind the lobe surface. Between the \HI tail region and the lobe surface no polarized signal is emitted, or if it is, that polarized signal is weaker than that emitted behind the \HI tail. We derived the \HI tail RM properties by simulating a RM over a path length of 11.04\,kpc, varying the B$_0 \cdot$n$_0$ parameter, and we add this term to the RM model of the control region. We obtain B$_0 \cdot$n$_0$=8.5$\cdot 10^{-3}$\,$\muup$G cm$^{-3}$, as shown in Figure \ref{fig:sdr_sum}.\\

We note that it is unlikely that a depolarization internal to the \HI tail is taking place, which would require  the \HI tail to be emitting a linearly polarized signal, which needs relativistic particles. These particles can be found in \HI regions if there is star formation. However, star formation is expected with column density higher than the values observed along the HI tail \citep[see][]{kleiner} and ionized gas should also be there, but we do not detect H$\alpha$ emission along the HI tail.
Therefore, we do not expect linear polarized emission from the HI tail itself nor internal depolarization.

\subsection{Density fluctuations in the \HI tail}
\begin{figure}
    \centering
    \includegraphics[width=0.49\textwidth]{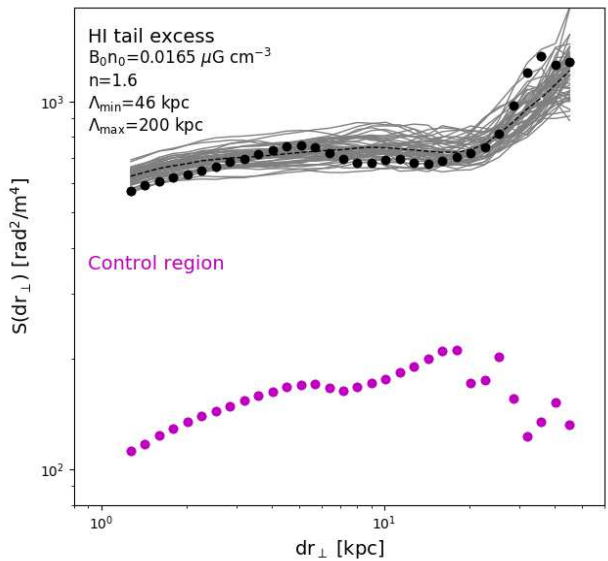}
    \caption{RM structure functions evaluated in the \HI tail (black) and in the control region (magenta). The gray lines near the \HI tail region are best-fit models for the \HI tail excess assuming that within the \HI tail the power spectrum of the thermal density is a power law with parameter reported on the left.}
    \label{fig:lls}
\end{figure}

In the previous section we modeled the RM structure function in the \HI tail region assuming that the corresponding Faraday screen volume is filled with a thermal plasma with constant density. This yielded a flat RM structure function (see  Figure \ref{fig:sdr_sum}), which cannot follow the rapid increment starting at $\sim$25\,kpc.
Here we try to relax the assumption of a constant density and, as we show later, we find a better match between the observed and the simulated RM structure functions.\\
To reach our goal, we assumed that the density has a power-law power spectrum fluctuating between a minimum and maximum scale. For the sake of simplicity, we assumed the same slope of the magnetic field power spectrum: n=1.6. We note that the results strongly depend on the minimum scale because with the chosen slope the energy is distributed on small scales. 
We reproduced the \HI tail excess as in the previous section: we created a Faraday screen with a magnetic field fluctuating according to the model in Section 4, and assuming a different average strength and different configuration of the thermal plasma density and structure; this corresponds to the \HI tail RM; the lobe contribution is modeled in the same way used in the previous section. The sum of the contributions is compared with the RM structure function to find the best-fit model of the thermal density power spectrum.
We explored several configurations with $\rm \Lambda_{min}$=0.6\,kpc and $\rm \Lambda_{max}$=2.4\,kpc to create a small-scale fluctuating structure: $\rm \Lambda_{min}$=0.6\,kpc and $\rm \Lambda_{max}$=46\,kpc; $\rm \Lambda_{min}$=26\,kpc, 46\,kpc, and 92\,kpc with $\rm \Lambda_{max}$=200\,kpc to simulate a structure fluctuating on large scales. \\
We obtained a simulation that shows an increment at large distances with $\rm \Lambda_{min}$=46\,kpc, $\rm \Lambda_{max}$=200\,kpc, and B$_0$n$_0$=16.5$\cdot 10^{-3}$\,$\muup$G cm$^{-3}$. 
Fifty realizations of this model are shown in Figure \ref{fig:lls} with thermal plasma density parameters reported in the top left corner.
We recall that the \HI tail analyzed in this work is the southern part of a larger structure, labeled T$_N$ in \citet{kleiner}, which extends for about 200\,kpc.

\subsection{HI tail embedded inside the Fornax\,A lobe, equilibrium conditions}
\begin{figure}
    \centering
    \includegraphics[width=0.49\textwidth]{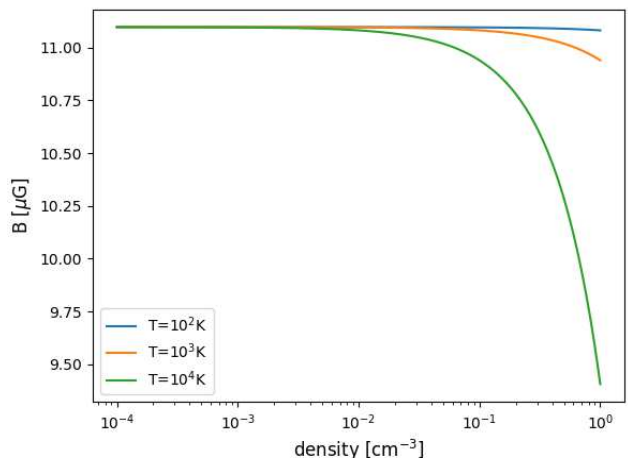}
    \caption{Magnetic field vs gas density according to the equilibrium condition of Eq. \ref{eq:equi} for three different values of the gas temperature.}
    \label{fig:equi}
\end{figure}

The exact location of the \HI tail along the line of sight cannot be determined with high accuracy.
However, as we already pointed out, since we can clearly associate the depolarizing effect with the \HI tail, this material should be either in the Fornax\,A lobe foreground or embedded in the lobe itself. Considering the large volume occupied by the lobe (approximately a sphere with a diameter of 25\arcsec$\sim$150\,kpc)  and the fact that the base of the tidal tail must be near the stellar body of NGC\,1316, we consider it more likely that the HI-detected part of the tail is within the lobe.\\
Here, we assumed that the \HI tail is within the Fornax\,A western lobe. This allowed us to obtain an estimate of the \HI tail magnetic field based on some physical considerations. \\
If the \HI tail is embedded in the lobe, we can reasonably assert that the external pressure exerted by the radio lobe ($\rm P_{nth}$) is balanced by the \HI tail pressure, which is due to the gas ($\rm P_{gas}$=nkT) and to the embedded magnetic field ($\rm P_B=$B$^2/8\pi$):
\begin{equation}
    \rm P_{nth}=P_{gas}+P_{B}.
    \label{eq:equi}
\end{equation}
Therefore, for a given nonthermal pressure and temperature, the magnetic field strength depends on the square root of the density.
Following this idea, we can derive an estimate of the magnetic field strength in the tail assuming a value for the radio lobe pressure from literature and typical values for the temperature and gas density. \citet{maccagni} determined the radio lobe pressure according to the equipartition condition, which gives a lower limit on this quantity ($\rm P_{nth} \geq 4.9 \cdot 10^{-11}$ erg cm$^{-3}$). The gas in the \HI tail should have a temperature between 10$^2$ and 10$^4$ K.
We show in Figure \ref{fig:equi} how the magnetic field strength varies with respect to the gas density in this range of temperatures.
Since the \HI tail is made of material originally belonging to a spiral galaxy, we expect to have typical gas densities on the order of 10$^{-2}$ cm$^{-3}$ in the \HI tail. Even assuming a denser gas, we estimated a magnetic field strength of around 9.5$-$11 $\muup$G, weakly affected by the gas temperature. It is worth noting that this value should be considered a lower limit since the radio lobe pressure was  determined under the equipartition condition.\\
Considering the \HI excess determined in Section 5.3, we derived a thermal plasma density of $\sim$8$-$9$\cdot 10^{-4}$\,cm$^{-3}$ in the \HI tail, or $\sim$1.5$-$1.7$\cdot10^{-3}$\,cm$^{-3}$ if the thermal gas is fluctuating as described in Section 5.4. In the next section we derive an upper limit on the thermal density from H$\alpha$ observations.

\section{H$\alpha$ constraints on thermal particles and magnetic field estimates}
We can tentatively give an estimate of the \HI tail thermal plasma density from H$\alpha$ observations using the Eq. 4 in Barger et al. 2013:
\begin{equation}
  \rm  L_{H+} = 2.75 \left ( \frac{T}{10^4 K} \right )^{0.924} \left ( \frac{I_{H\alpha}}{R} \right )\left ( \frac{n_e}{cm^{-3}} \right )^{-2} \rm pc.
\end{equation}
The dimension of the cloud $\rm L_{H+}$ can be determined from the width of the \HI tail, under the assumption of a cylindrical structure. This gives $\rm L_{H+}$ = 11.04 kpc. \citet{kleiner} did not detect H$\alpha$ along the tail and we can use a 3$\rm\sigma$ threshold ($\rm \sigma = 2 \cdot 10^{-18} erg \cdot cm^{-2} \cdot s^{-1} \cdot arcsec^{-2}$) of their observations as an upper limit for the H$\rm \alpha$ brightness in this region, I$\rm _{H\alpha}$. \\
In a reasonable range of gas temperatures for an H$\alpha$ cloud ($10^3-10^4$ K) the thermal electron density is between 2$\cdot 10^{-3}$ and 6$\cdot 10^{-3}$\,cm$^{-3}$. This means that the thermal electron density in the \HI tail is below 6$\cdot 10^{-3}$\,cm$^{-3}$. \\
This value is compatible with the findings of the previous sections: if the \HI tail is embedded in the Fornax\,A western lobe, it is hosting a magnetic field with a strength of $\sim$9.5$-$11\,$\muup$G and a thermal plasma density of $\sim$8$-$9$\cdot 10^{-4}$\,cm$^{-3}$ constant within the tail, or $\sim$1.5$-$1.7$\cdot 10^{-3}$\,cm$^{-3}$ if it fluctuates. In both cases the thermal plasma density is below the detection threshold of the H$\alpha$ observations by \citet{kleiner}. 

\section{Summary and conclusions}
In this work we analyzed the spatial coincidence between \HI tidal material and a depolarized structure in the western lobe of Fornax\,A. Our goal was to understand if the \HI tail is acting as depolarizing Faraday screen. Our results are summarized in Table \ref{tab:report} and discussed in what follows.\\
\begin{table*}[ht]
\renewcommand*{\arraystretch}{1.4}
\centering
\begin{tabular}{ l  c c c c c c c} 
 \hline \hline
 \multicolumn{4}{l}{{\bf Faraday screen from the Fornax cluster intracluster medium}} \\
 \hline
 Region & n$_e$(r) & n$_e$ power-law & B(r) & B power-law & L & B$_0$ & Plausible \\ 
 & & power spectrum & & power spectrum & [kpc] & [$\muup$G] &\\
 \hline \hline
 Control region & $\beta-$model & no & $\eta-$model & n=2 & 1000 & 900 & no \\ 
 & \citep{chen} & & with $\eta$=0.5 & & & \\ 
 \HI tail region & $\beta-$model & no & $\eta-$model & n=1.6 & 1000 & 2000 & no \\ 
 & \citep{chen} & & with $\eta$=0.5 & & & \\ 
 \hline \hline
 \multicolumn{4}{l}{{\bf Faraday screen from the Fornax\,A intra-group}} \\
 \hline 
 Region & n$_e$(r) & n$_e$ power-law & B(r) & B power-law & L & B$_0$ & Plausible \\ 
 & & power spectrum & & power spectrum & [kpc] & [$\muup$G] &\\
 \hline \hline
 Control region & $\beta-$model & no & $\eta-$model & n=2 & 1000 & 18.5 & yes \\
  & \citep{babyk} & & with $\eta$=0.5 && & \\ 
 \HI tail region & $\beta-$model& no & $\eta-$model & n=1.6 & 1000 & 50 & no \\
  & \citep{babyk} & & with $\eta$=0.5 & & & \\ 
 \hline \hline
 \multicolumn{4}{l}{{\bf Faraday screen assuming the advection scenario}} \\
 \hline
 Region & n$_e$(r) & n$_e$ power-law & B(r) & B power-law & L & B$_0$n$_0$ & Plausible \\ 
 & & power spectrum & & power spectrum & [kpc] & [$\muup$G\,cm$^{-3}$] &\\
 \hline \hline
 Control region & constant & no & constant & n=2 & 11.04 & 3.5$\cdot 10^{-3}$ & yes \\
 \HI tail region & constant & no & constant & n=1.6 & 11.04 & 9.0$\cdot 10^{-3}$ & yes \\
 \HI tail excess & constant & no & constant & n=1.6 & 11.04 & 8.5$\cdot 10^{-3}$ & yes \\
 \HI tail excess & constant & n=1.6 & constant & n=1.6 & 11.04 & 16.5$\cdot 10^{-3}$ & yes \\
 & & $\rm \Lambda_{min}=46$\,kpc & & & & &\\
 & & $\rm \Lambda_{min}=200$\,kpc & & & & &\\
 \hline \hline
 \multicolumn{4}{l}{\bf{Pressure equilibrium}} \\
 \hline
 Region & Gas density & Temperature & & Nonthermal pressure & & B$_0$ & Plausible\\
 & [cm$^{-3}$] & [K] & & [erg cm$^{-3}$] & & [$\muup$G] &\\
 \hline  \hline
 \HI tail & 10$^{-2}$ & 10$^2-10^4$ & & 4.9 $\cdot 10^{-11}$ & & 9.5$-$11 & yes\\
 \hline
 \\
\end{tabular}
\caption{Summary of the results. We specify the scales of fluctuation of the density power spectra since they are different with respect to those of the magnetic field (see   text for further details).}
\label{tab:report}
\end{table*}

We modeled the magnetic field power spectra along the \HI tail and in a control region with the same dimension and orientation by making a 2D simulation of the RM structure function. Keeping the minimum and maximum scale of fluctuation fixed to $\rm\Lambda_{min}$=0.6\,kpc and $\rm\Lambda_{max}$=46\,kpc, we found power spectra with different normalizations and slopes. In particular, the control region showed a magnetic field power spectrum with slope n=2, normalization equal to 75\,$\rm\muup$G Mpc$^{-3}$, and a RM auto-correlation length of $\rm\Lambda_{RM}$=3.3\,kpc, while the \HI tail region has a slope of n=1.6, a normalization of 350\,$\rm\muup$G Mpc$^{-3}$, and a RM auto-correlation length of $\rm\Lambda_{RM}$=1.9\,kpc.  \\
We then used an analytical approach and 3D simulations to understand what physical scenario is producing the observed RM structures. We found that it is unlikely that the observed RM is due to the intracluster medium of the Fornax cluster since we would need a very strong magnetic field  (i.e., on the order of thousands of $\muup$G). A modeling of the intragroup medium of the Fornax\,A group could explain the RM observed in the control region if the central magnetic field strength were 18.5\,$\rm\muup$G. However, such modeling cannot explain simultaneously the RM of the control region and that observed at the \HI tail location. Therefore, it seems clear that the \HI tail itself is acting as a Faraday screen. \\
We also investigated  the possibility that the RM in the control region is due to the advection of magnetized plasma on the lobe surface of Fornax\,A, as suggested by \citet{anderson18b}. Assuming a magnetic field fluctuating  around an average strength (as described in Section 4) and a constant and uniform thermal density, and by integrating the Faraday screen over a length of 11.04\,kpc (i.e., the \HI tail width), we found B$_0 \cdot$n$_0$=3.5$\cdot 10^{-3}$\,$\muup$G cm$^{-3}$ in the control region. For the \HI tail, we considered two scenarios,  first where the RM is entirely due to the \HI tail, which gives an upper limit of  B$_0 \cdot$n$_0$=9$\cdot 10^{-3}$\,$\muup$G cm$^{-3}$, and  second where the RM is due to the sum of the lobe RM and  the \HI tail RM. The second scenario applies if the \HI tail is in the lobe foreground or within the lobe, very close to its surface. In this case, we obtained B$_0 \cdot$n$_0$=8.5$\cdot 10^{-3}$\,$\muup$G cm$^{-3}$ in the \HI tail region.
We noted that the rapid increment of the \HI tail region RM structure function at large spatial scales cannot be traced by this modeling. Therefore, we assumed that the thermal plasma has a power-law power spectrum with index n=1.6. We found that only a thermal plasma fluctuating between a minimum and a maximum scale of 46\,kpc and 200\,kpc, respectively, can reproduce the increase of the structure function at distances larger than 25\,kpc. In this case the parameter B$_0\cdot$n$_0$=16.5$\cdot 10^{-3}$\,$\muup$G cm$^{-3}$.
We noted that there is degeneracy between the minimum scale of fluctuation and the slope of the thermal plasma power spectrum. We do not have further constraints on the thermal plasma distribution in the \HI tail since H$\alpha$ is not detected at this location or in the lobe surface. An exact reconstruction of the distribution of the thermal density is beyond the scope of this paper.\\
Following the idea that the \HI tail is embedded in the Fornax\,A lobe, we derived an estimate of the magnetic field in the \HI tail considering the equilibrium condition where the nonthermal pressure exerted by the radio lobe is balanced by the \HI tail pressure, which is due to the gas and to the magnetic field. From this equation we obtained a magnetic field strength of $\sim$9.5$-$11\,$\muup$G for a gas temperature between 10$^2$ and 10$^4$\,K. These values represent an upper limit since the nonthermal pressure of the radio lobes was   derived under equipartition condition. \\
We used H$\alpha$ observations to find an upper limit on the thermal plasma distribution in the \HI tail and in the control region. The result is a thermal plasma density below 6$\cdot 10^{-3}$\,cm$^{-3}$.\\
From the 3D simulation of the RM structure function associated with the \HI tail, we obtained a magnetic field strength times the thermal plasma distribution of B$_0 \cdot$n$_0$=8.5$\cdot 10^{-3}$\,$\muup$G cm$^{-3}$ assuming a constant thermal density, and B$_0 \cdot$n$_0$=16.5$\cdot 10^{-3}$\,$\muup$G cm$^{-3}$ if the thermal density is fluctuating. From the equilibrium condition we know that the \HI tail host a magnetic field of $\sim$9.5$-$11\,$\muup$G. Combining all these findings we obtained a thermal plasma distribution of $\sim$8$-$9$\cdot10^{-4}$\,cm$^{-3}$ or $\sim$1.5$-$1.8$\cdot 10^{-3}$\,cm$^{-3}$\,cm$^{-3}$ for a constant and a fluctuating thermal density, respectively. Both values are compatible with the upper limit from H$\alpha$ observations.\\

From this analysis is clear that we have found an additional Faraday screen due to the \HI tail. 
We therefore conclude that the tail, which is the remnant of a galaxy merger that occurred $\sim2$ Gyr ago and is made of material originally belonging to a Milky Way-like galaxy, is either driving a multi-phase magnetized gaseous medium through the radio lobe of Fornax A or that it survived the lobe expansion.
This could be the first observed evidence of such phenomena.

\begin{acknowledgements}
This project has received funding from the European Research Council (ERC) under the European Union’s Horizon 2020 research and innovation programme (grant agreement no. 679627; project name FORNAX). 
FL acknowledges financial support from the Italian Minister for Research and Education (MIUR), project FARE, project code R16PR59747, project name FORNAX-B. 
FL acknowledges financial support from the Italian Ministry of University and Research $-$ Project Proposal CIR01$\_$00010.
VV and MM acknowledge support from INAF mainstream project "Galaxy Clusters Science with LOFAR" 1.05.01.86.05.
Part of the data published here have been reduced using the CARACal pipeline, partially supported by ERC Starting grant number 679627 "FORNAX", MAECI Grant Number ZA18GR02, Department of Science and Innovation$-$NRF Grant Number 113121 as part of the ISARP Joint Research Scheme, and BMBF project 05A17PC2 for D$-$MeerKAT. Information about CARACal can be obtained online under the URL: https://caracal.readthedocs.io.
We also thank our anonymous referee for their constructive feedback which has strengthened this work.
\end{acknowledgements}

%
%

\end{document}